\documentclass{pasj01}
\usepackage{graphicx}
\Received{$\langle$reception date$\rangle$}
\Accepted{$\langle$acception date$\rangle$}
\Published{$\langle$publication date$\rangle$}
\usepackage[switch,mathlines]{lineno}
\usepackage{xspace}
\usepackage{color}
\usepackage{bm}
\usepackage[super]{nth}

\DeclareGraphicsExtensions{.pdf,.png,.jpg}

\begin{document}
\newcommand{\km}{\>$\mathrm{km\>s^{-1}}$}
\newcommand{\ngc}{NGC\,7738\xspace}
\newcommand{\hto}{\atom{H_2O}{}{}\xspace}
\newcommand{\vlsr}{$V_\mathrm{LSR}$}
\newcommand{\vrot}{$V_\mathrm{rot}$}
\newcommand{\ain}{$a_\mathrm{in}$}
\newcommand{\aout}{$a_\mathrm{out}$}
\newcommand{\rin}{$r_\mathrm{in}$}
\newcommand{\rout}{$r_\mathrm{out}$}
\newcommand{\mbh}{$M_\mathrm{BH}$}
\newcommand{\mdisk}{$M_\mathrm{disk}$}
\newcommand{\mc}{$M_\mathrm{c}$}
\newcommand{\red}[1]{\textcolor{red}{#1}}


\title{Water Maser Disk and a Supermassive Black Hole at the Nucleus of the Active Galaxy NGC\,7738}
\author{Rinka \textsc{Ito},\altaffilmark{1,}$^{*}$
        Yusuke \textsc{Miyamoto},\altaffilmark{2,}$^{*}$
        Naomasa \textsc{Nakai},\altaffilmark{1,}$^{*}$
        Aya \textsc{Yamauchi}\altaffilmark{3,}$^{*}$
        and
        Yuichi \textsc{Terashima}\altaffilmark{4,}$^{*}$
        }%
\altaffiltext{1}{Department of Physics, School of Science and Technology, Kwansei Gakuin University, 1 Gakuen Uegahara, Sanda, Hyogo 669-1330}
\altaffiltext{2}{{Department of Electrical, Electronic and Computer Engineering, Faculty of Engineering, Fukui University of Technology, 3-6-1, Gakuen, Fukui 910-8505}}
\altaffiltext{3}{Mizusawa VLBI Observatory, National Astronomical Observatory of Japan, 2-12 Hoshigaoka, Mizusawa, Oshu, Iwate 023-0861}
\altaffiltext{4}{Department of Physics, Faculty of Science, Ehime University, Matsuyama, Ehime 790-8577}

\email{dbc11115@yahoo.co.jp,
      {yusuke.miyamot@gmail.com},
      ilo7771@kwansei.ac.jp,
      a.yamauchi@nao.ac.jp,
      terashima.yuichi.mc@ehime-u.ac.jp}


\maketitle

\begin{abstract}
We present the results of VLBI observations of water vapor masers in the nucleus of the LINER galaxy \ngc.
The red- and blue-shifted and newly detected systemic maser features show an almost edge-on disk located at a distance of ${0.031}\mbox{--}{0.222}$\>pc from the galactic center and rotating with a velocity of ${324}\mbox{--}{454}$\km.
The velocity field of the disk indicates sub-Keplerian rotation, suggesting a non-negligible disk mass.
The Mestel disk model reveals the central and disk masses to be $({1.2}\,\pm\,{0.4})\times10^6 \Mo$ and $({{4.7}\,\pm\,{1.5}})\times{10^{6}} \Mo$, respectively.
The mean volume density within the inner radius of the disk  [$({1.2}\,\pm\,{0.5})\times{10^{10}}\Mo\>\mathrm{pc^{-3}}$] strongly suggests the existence of a supermassive black hole at the center.
\end{abstract}

\section{Introduction}
Water vapor (\hto) maser emission directly probes the structure and dynamics of active galactic nuclei (AGN) at the (sub-)parsec scale.
With its extremely high brightness temperature  ($T_\mathrm{B}\,\sim\,10^{10}\>\mathrm{K}$), it is one of few emission lines observable with very long baseline interferometry (VLBI).
VLBI allows observations of dense gas distribution and motion in nuclei at milliarcseconds (mas).
VLBI observations of \hto maser emission in the active galaxy NGC\,4258 revealed a compact disk in Keplerian rotation and a massive black hole in its nucleus
(\cite{Miyoshi1995}).
This paper presents the results of VLBI observations of the \hto maser in \ngc.

\begin{figure}
\begin{center}
   \includegraphics[width=80mm]{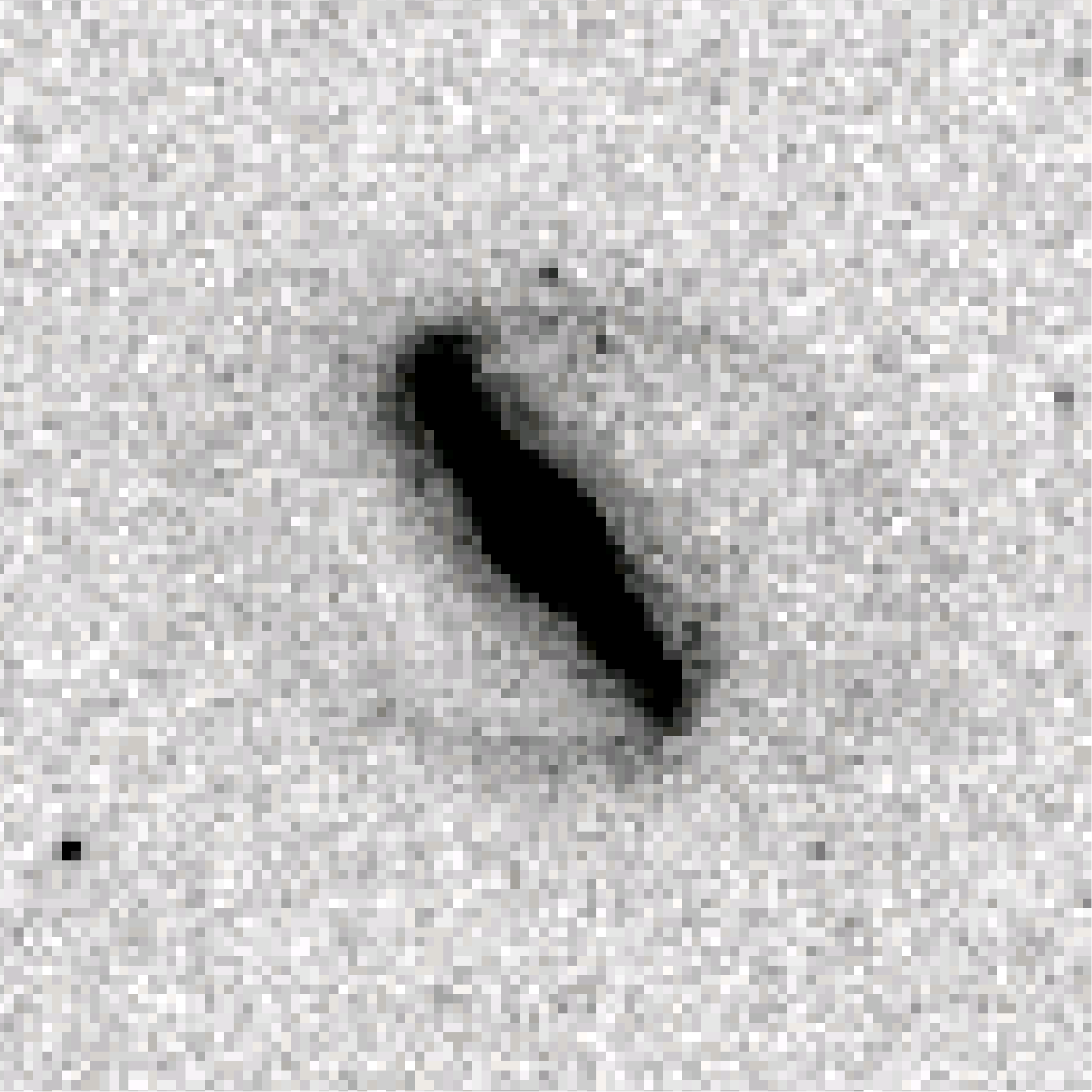}
\end{center}
\caption{
  \textrm{Optical image of NGC\,7738 adopted from Digitized Sky Surveys.\footnotemark
 The scale is \timeform{3'.0}\,$\times$\,\timeform{3'.0}
 in R.A. and Decl., and north is top and east is left.
  }}
\end{figure}\label{fig:opt}
\footnotetext{The Digitized Sky Surveys were produced at the Space Telescope Science Institute under the U.S. Government grant NAG W-2166.
Images from these surveys are based on photographic data obtained using the Oschin Schmidt Telescope on Palomar Mountain and the UK Schmidt Telescope.}

\ngc (UGC\,12757) is a nearly face-on galaxy classified as SBb
(\cite{Nilson1973}; \cite{Walcher2014});
its optical image is shown in figure \ref{fig:opt}.
While the optical spectrum is classified as an H\, \emissiontype{II} nucleus (\cite{Kewley2001}),
the nucleus is an obscured AGN,
that is, the X-ray spectrum shows flat continuum spectra and a strong Fe--K line, indicating the presence of a Compton-thick AGN
(\cite{Terashima2015}).

\citet{Yamauchi2017} discovered \hto maser emission from \ngc.
The maser spectrum showed red- and blue-shifted features and possibly weak systemic velocity features.
The shifted features offset from the systemic velocity by $\Delta V=331-371\>\mathrm{km\>s^{-1}}$.
The total isotropic luminosity was $L=468\>\LO$,
indicating a megamaser.
We report the detection of all maser features using VLBI,
the nuclear structure of the galaxy, and the mass of a black hole separated from  {the mass} of a disk rotating around the black hole.
We adopt velocities in the radio definition and with respect to the local standard of rest (LSR).
The heliocentric velocity is
$V_\mathrm{{hel}}=V_\mathrm{{LSR}}+0.2$\km{}
for this galaxy.
Basic parameters of \ngc are listed in table \ref{tab:NGC7738},
where we adopted the angular-size distance of $88.9\,\pm\,3.6\>\mathrm{Mpc}$,
using a flat $\Lambda$CDM cosmology,
the Hubble constant $H_0=73.9\,\pm\,3.0\>\mathrm{km\>s^{-1}\>Mpc^{-1}}$, and a density parameter $\Omega_m=0.315$.
{One mas corresponds to 0.43\>pc at the angular-size distance.}


 \begin{table}
 \begin{center}
  \tbl{Adopted parameters of NGC\,7738.}{%
  \begin{tabular}{@{}ll@{}}
  \hline\noalign{\vskip3pt}
    Center position\footnotemark[$*$] & $\alpha_{2000}=$ \timeform{23h44m2s.029} \\
     & $\delta_{2000}=$ +\timeform{0D30'59''.97} \\
     Systemic velocity\footnotemark[$*$] & $V_\mathrm{LSR, rad}={6598}\,\pm\,5$\km \\
      & $V_\mathrm{LSR, opt}=cz={6746\,\pm\,5}$\km \\
     Morphological type\footnotemark[$\dag$] & SBb \\
   Angular-size distance\footnotemark[$\ddag$] & $88.9\,\pm\,3.6\>\mathrm{Mpc}$ \\
   Luminosity distance & $92.9\,\pm\,3.8\>\mathrm{Mpc}$ \\
   Inclination angle\footnotemark[$\dag$] & $\sim\timeform{0D}$ \\
  \hline\noalign{\vskip3pt}
 \end{tabular}}\label{tab:NGC7738}
 \begin{tabnote}
   \par\noindent
   \footnotemark[$*$] This paper.
   \par\noindent
   \footnotemark[$\dag$] \cite{Nilson1973} ;\cite{Walcher2014}.
   \par\noindent
   \hangindent6pt\noindent
   \hbox to6pt{\footnotemark[$\ddag$]\hss}\unskip%
     Angular-size distance $D$ is calculated,
     using
     $D\approx{cz}/{(H_0(1+z))}(1-3\Omega_mz/4+\Omega_m(9\Omega_m-4)z^2/8)$
     (\cite{Pesce2020}),
     where adopted $H_0=73.9\,\pm\,3.0\>\mathrm{km\>s^{-1}\>Mpc^{-1}}$ and $\Omega_m=0.315$ (see text).
 \end{tabnote}
 \end{center}
 \end{table}
\section{Observations}

VLBI observations of \atom{H_2O}{}{} maser emission
($J_\mathrm{K_aK_c}=6_{16}-5_{23}$ transition at the rest frequency of 22.23508\>GHz)
were made using the Very Long Baseline Array (VLBA)
with the Very Large Array (VLA) of the National Radio Astronomy Observatory\footnote{
The National Radio Astronomy Observatory is a facility of the National Science Foundation, US, operated under cooperative agreement by Associated Universities, Inc.}
(NRAO US) on 2019 January 25 and 26, for 5 h each day{; the} VLA was operated as a phased array.
Two IFs were recorded
{in left and right circular polarization (LCP, RCP), respectively,}
each with a bandwidth of 128\>MHz (1765\km) divided into 2048 channels (0.86\km{} per channel).
The LSR {velocity} at the band center of each IFs {was} 6580\km{, and thus, the observed velocity range was 5698--7462\km}.
During observations on Day 25th, one antenna of the VLBA at Kitt Peak (KP) was  unavailable.

The data were processed on the VLBA correlator at NRAO, and after correlation, data reduction, including calibration and imaging, were processed with the Astronomical Image Processing System (AIPS) package.
3C\,454.4 was used as a fringe finder{, and} 2335--027 and 2351--006 as phase reference sources.
{We obtained phase-only self-calibration solutions for the strangest maser features at \vlsr=6580.5\km{} and applied them to the continuum visibilities.
LCP and RCP were averaged.
The imaging was performed with the CLEAN-method,
using UNIFORM weighting.}
The maser features were {averaged over} 5 channels {each} (3.37\km) {to reduce noise}.

To image the continuum emission, we averaged over the channels at velocity ranges of 5842--6469\km{} and 7106--7316\km, avoiding maser features.
The synthesized beam sizes and image noise for both the line and continuum images are listed in table \ref{tab:rms}.

\begin{table}
\begin{center}
 \tbl{Beam size and rms noise of VLBI image.}{%
 \begin{tabular}{@{}lccccc@{}}
 \hline\noalign{\vskip3pt}
 & Channels & Major & Minor & PA & Rms noise \\
 & averaged & (mas) & (mas) & (\timeform{D}) & ($\mathrm{mJy}\>\mathrm{beam^{-1}}$) \\
 \hline\noalign{\vskip3pt}
 Spectrum line & 5 & 1.36 & 0.48 & -10 & 0.59 \\
 continuum & & 0.95 & 0.36 & -8 & 0.04 \\
  \hline\noalign{\vskip3pt}
\end{tabular}}\label{tab:rms}
 \end{center}
\end{table}

\section{Results}
\subsection{Detection of maser emission}

\begin{figure}
\begin{center}
  \includegraphics[width=80mm]{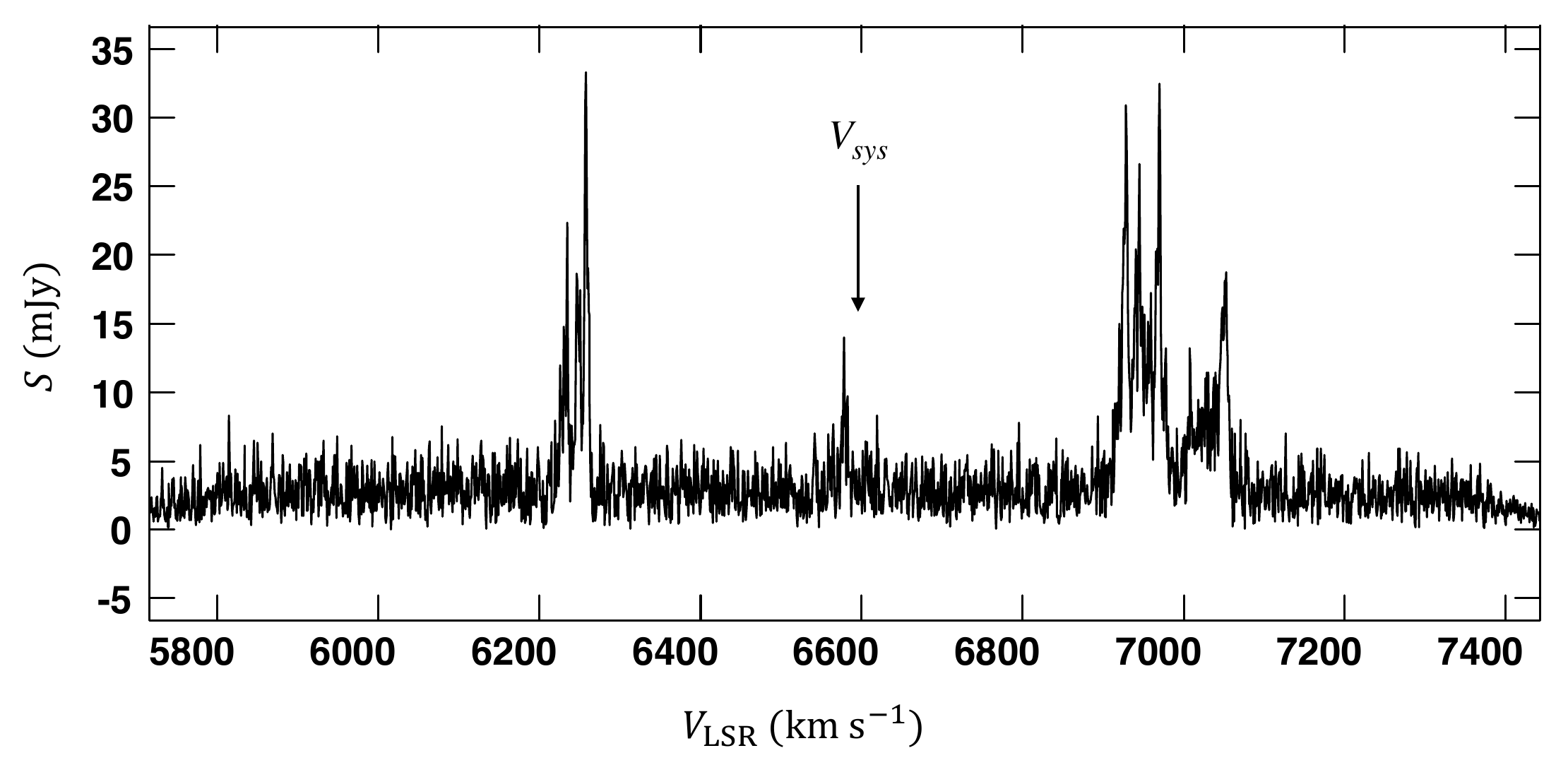}
\end{center}
  \caption{
  \textrm{
  {Cross-power spectral amplitudes of the visibilities with four antennas (FD, LA, PT, and VLA) toward \ngc;
  we excluded longer-baseline visibilities for this spectrum to avoid coherence loss.
  No continuum $(2.7\>\mathrm{mJy})$ subtraction is applied.}
  The arrow denotes the systemic velocity of the galaxy, $V_\mathrm{LSR}={6598}$\km.
  The newly detected maser features are observed at $V_\mathrm{LSR}=6559$--$6613$\km (systemic features).
  }
  }
  \label{fig:spectrum}
\end{figure}


Figure \ref{fig:spectrum} shows the spectrum obtained by concatenating the spectra over all observation days.
Red- and blue-shifted features are observed at
$V_\mathrm{LSR}={6902}\mbox{--}{7065}$\km{} and
$V_\mathrm{LSR}= {6227}\mbox{--}{6264}$\km,
respectively,
consistent with a previous report (\cite{Yamauchi2017}).
In addition to these features,
systemic features are visible at $V_\mathrm{LSR}=6559$--6613\km{} (also see \cite{Yamauchi2017}).
The red- and blue-shifted features with respect to the system velocity ($V_\mathrm{sys}={6598}$\km)
are almost symmetric:
$|V_\mathrm{red}-V_\mathrm{sys}|={305}\mbox{--}{468}$\km{}
and $|V_\mathrm{blue}-V_\mathrm{sys}|={333}\mbox{--}{370}$\km.
The apparent isotropic luminosity of the \hto maser are $L=${115}\LO, {36}\LO, and {5}\LO
for the red-shifted, blue-shifted, and systemic features,
respectively, at a luminosity distance of 92.9\>Mpc.

\begin{figure*}[h]
  \begin{center}
    \includegraphics[width=160mm]{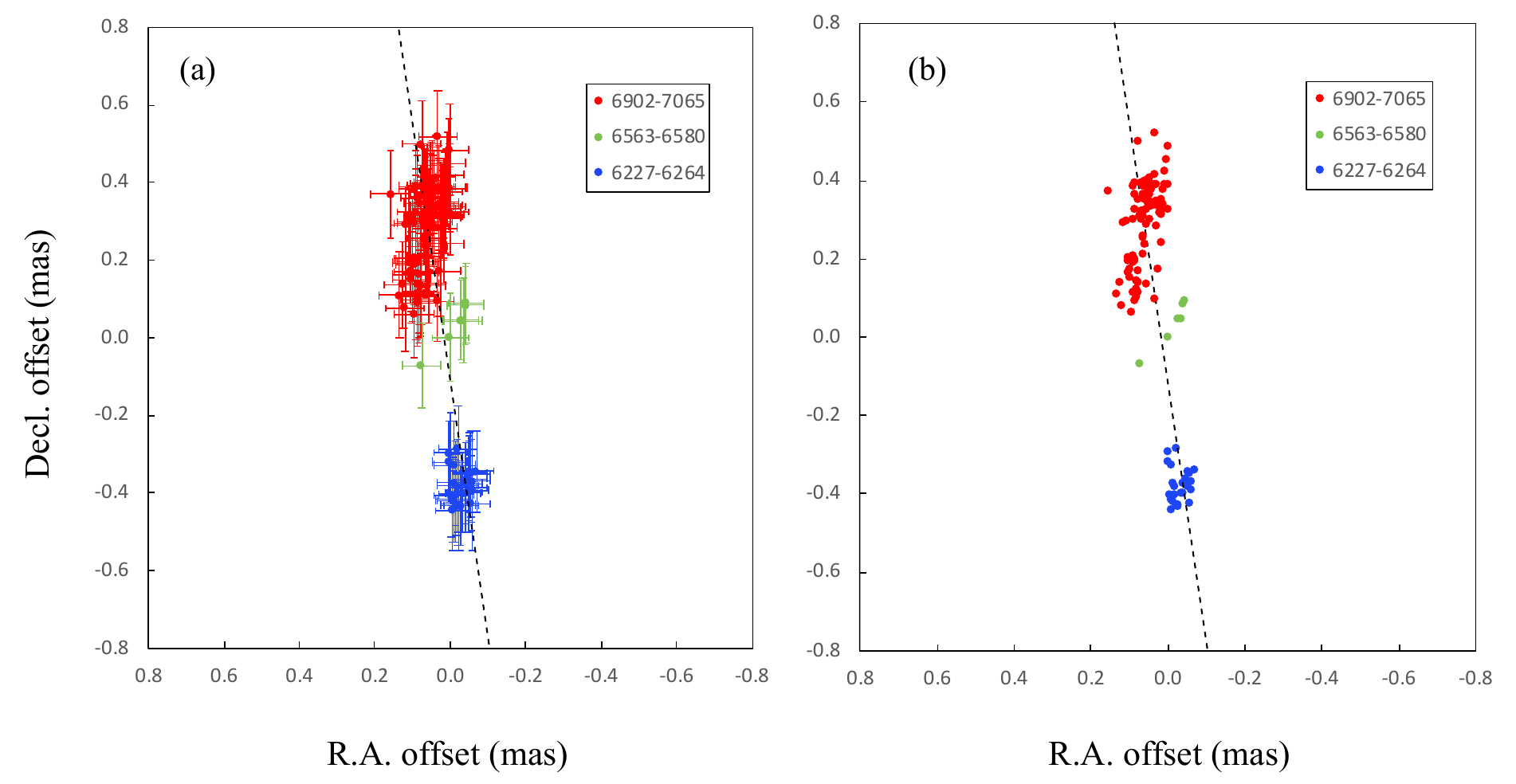}
  \end{center}
  \caption{
  \textrm{
  (a) Distribution of the maser spots in the nuclear region of NGC\,7738.
  (b) Same as (a) but without error bars.
  The origin of the coordinates, ($\Delta \mathrm{RA}, \Delta \mathrm{Decl})=(0\>\mathrm{mas}, 0\>\mathrm{mas})$,
  is the peak position of the strongest maser feature of the systemic features at $V_\mathrm{LSR}=6581$\km.
  The colored circles indicate the peak positions of the maser features
  detected at the $\geq {6}\>\sigma$ level.
  The color indicates the LSR velocity of the maser spots in \km.
  Dotted lines with a position angle of $\mathrm{PA}={\timeform{8.6D}}$ are the result of a least-squares fitting of the circles with errors in (a){, being not through $(\Delta\mathrm{RA}, \Delta \mathrm{Decl})=(0\>\mathrm{mas},0\>\mathrm{mas})$}.
  }
}
  \label{fig:spots}
\end{figure*}

\subsection{Distribution of maser emission}
Figure \ref{fig:spots} shows the distribution of maser emission detected at  $\geq {6}\>\sigma$ level,
where
$1\>\sigma=0.59\>\mathrm{mJy\>beam^{-1}}$,
and table \ref{tab:posi} {lists} the maser spots positions (all data in the e-table.\footnote{E-table 1 is available only on the online edition as supplementary data}).
The peak position of the strongest maser feature ($5\>\mathrm{mJy\>beam^{-1}}$) of  systemic features at $V_\mathrm{LSR}=6581$\km{} was adopted
as the coordinate origin.
Position errors were
$\Delta\theta(\mathrm{rms})=0.5\theta_\mathrm{beam}/\mathrm{SNR}=${0.05--0.14 and 0.02--0.06\>mas} in major and minor axis directions of the synthesized beam, respectively,
where $\theta_\mathrm{beam}$ is synthesized beam size, and SNR is the signal-to-noise ratio of peak emission.
Maser spots aligned along a line at $\mathrm{PA}={\timeform{8.6D}}\,\pm\,{\timeform{0.8D}}$, determined by least-squares fitting.
Positions of the strongest flux densities for the red-shifted, blue-shifted, and systemic features are
($\Delta \mathrm{RA}, \Delta \mathrm{Decl})=(0.05\>\mathrm{mas}, 0.39\>\mathrm{mas})$,
$(-0.02\>\mathrm{mas}, -0.43\>\mathrm{mas})$, and
$(0\>\mathrm{mas}, 0\>\mathrm{mas})$, respectively.

{Continuum emission was not detected
in the map, because the continuum was probably resolved out by the long baselines.
The upper limit was 3\,$\sigma=0.13\>\mathrm{mJy\>beam^{-1}}$.}

\begin{table*}
\begin{center}
 \tbl{Positions of the maser spots.\footnotemark[$*$]}{%
 \begin{tabular}{@{}lcccccc@{}}
 \hline\noalign{\vskip3pt}
  & $V_\mathrm{LSR}$ & Flux Density & RA Offset & Decl Offset & $X$\footnotemark[$\dag$] & $Y$\footnotemark[$\dag$] \\
  & ($\mathrm{km\>s^{-1}}$) & ($\mathrm{mJy}\>\mathrm{beam^{-1}}$) & (mas) & (mas) & (mas) & (mas) \\
 \hline\noalign{\vskip3pt}
  Red-shifted
  &	6926.8	&	9.1	\,$\pm$\,	1.0	&	0.048	\,$\pm$\,	0.044	&	0.39	\,$\pm$\,	0.10	&	0.52	\,$\pm$\,	0.07	&	{0.028}	\,$\pm$\,	0.028	\\
  &	6926.0	&	8.5	\,$\pm$\,	0.9	&	0.034	\,$\pm$\,	0.043	&	0.39	\,$\pm$\,	0.10	&	{0.50}	\,$\pm$\,	0.06	&	{0.041}	\,$\pm$\,	0.028	\\
  &	6925.1	&	7.6	\,$\pm$\,	0.8	&	0.014	\,$\pm$\,	0.044	&	0.38	\,$\pm$\,	0.10	&	{0.49}	\,$\pm$\,	0.07	&	{0.059}	\,$\pm$\,	0.029	\\
  \hline\noalign{\vskip3pt}
  Systemic
  &	6580.5	&	5.1	\,$\pm$\,	0.7	&	0.000	\,$\pm$\,	0.048	&	0.00	\,$\pm$\,	0.11	&	{0.11}	\,$\pm$\,	0.09	&	{0.017}	\,$\pm$\,	0.035	\\
  &	{6579.6}	&	{3.8}	\,$\pm$\,	{0.5}	&	{-0.042} \,$\pm$\, {0.047}	&	{0.09}	\,$\pm$\,	{0.10}	&	{0.20} \,$\pm$\,	{0.07}	&	{0.072}	\,$\pm$\,	{0.034}	\\
  &	{6578.8}	&	{4.1}	\,$\pm$\,	{0.6}	&	{-0.040}	\,$\pm$\,	{0.047}	&	{0.08}	\,$\pm$\,	{0.10}	&	{0.19}	\,$\pm$\,	{0.07}	&	{0.069}	\,$\pm$\,	{0.033}	\\
  \hline\noalign{\vskip3pt}
  Blue-shifted
  &	6259.4	&	9.2	\,$\pm$\,	1.2	&	-0.008	\,$\pm$\,	0.047	&	-0.42	\,$\pm$\,	0.11	&	{-0.30}	\,$\pm$\,	0.08	&	{-0.038}	\,$\pm$\,	0.033	\\
  &	6258.5	&	10.6	\,$\pm$\,	1.3	&	-0.014	\,$\pm$\,	0.046	&	-0.42	\,$\pm$\,	0.11	&	{-0.31}	\,$\pm$\,	0.08	&	{-0.032}	\,$\pm$\,	0.033	\\
  &	6257.7	&	11.2	\,$\pm$\,	1.4	&	-0.022	\,$\pm$\,	0.046	&	-0.43	\,$\pm$\,	0.11	&	{-0.32}	\,$\pm$\,	0.08	&	{-0.025}	\,$\pm$\,	0.032	\\
 \hline\noalign{\vskip3pt}
\end{tabular}}\label{tab:posi}
\begin{tabnote}
  \par\noindent
  \footnotemark[$*$] This is just a sample. All data are available online.
  \par\noindent
  \footnotemark[$\dag$] $X$ and $Y$ are along the lines of $\mathrm{PA}={\timeform{8.6D}}$ and ${\timeform{-81.4D}}$, respectively.
\end{tabnote}
\end{center}
\end{table*}


\section{Discussion}
\subsection{Rotating disk}
The maser spots of \ngc were mostly distributed along the line,
$\mathrm{PA}={\timeform{8.6D}}$ (figure \ref{fig:spots}):
The red-shifted features were located on the northeastern side, and the blue-shifted features were located on the southwestern side of the systemics velocity.
Based on the distribution, {a} masing disk is considered, being an edge-on disk with a position angle of $\mathrm{PA}={\timeform{8.6D}}$.
The maser disk is almost perpendicular to the kpc-scale galactic disk which is face-on
(figure \ref{fig:opt} and table \ref{tab:NGC7738}).
Such misalignment has often been observed in megamasers
(e.g.,
\cite{Miyoshi1995};
\cite{Braatz1997};
\cite{Greenhill2009};
\cite{Yamauchi2012}
).

\begin{figure*}
  \begin{center}
    \includegraphics[width=160mm]{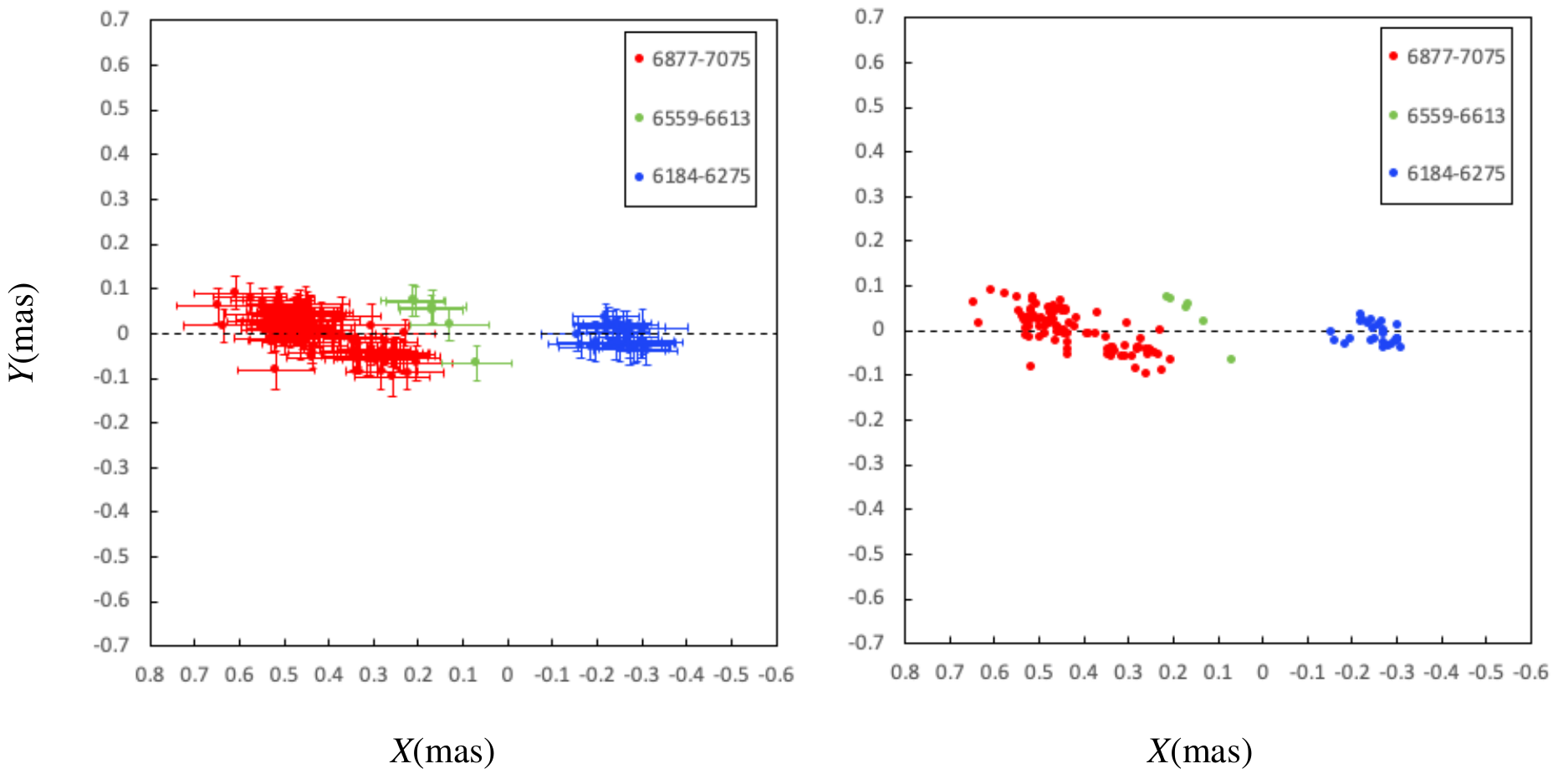}
  \end{center}
  \caption{\textrm{
  (a) Distribution of maser spots in the frame of the coordinates of $X$ along $\mathrm{PA}={\timeform{8.6D}}$.
  (b) Same as (a) but without error bars.
  The figures are a rewrite of figure \ref{fig:spots} with the dotted line ($\mathrm{PA}={\timeform{8.6D}}$) in figure \ref{fig:spots} as the $X$ line.
  The origin of the coordinates,
  ($\Delta{X}, \Delta Y)=(0\>\mathrm{mas}, 0\>\mathrm{mas})$,
  is equal to ($\Delta \mathrm{RA}, \Delta \mathrm{Decl})=(0.00\>\mathrm{mas}, 0.13\>\mathrm{mas})$
  in figure \ref{fig:spots}.
  }
  }
  \label{fig:pa}
\end{figure*}

\begin{figure*}
  \begin{center}
    \includegraphics[width=160mm]{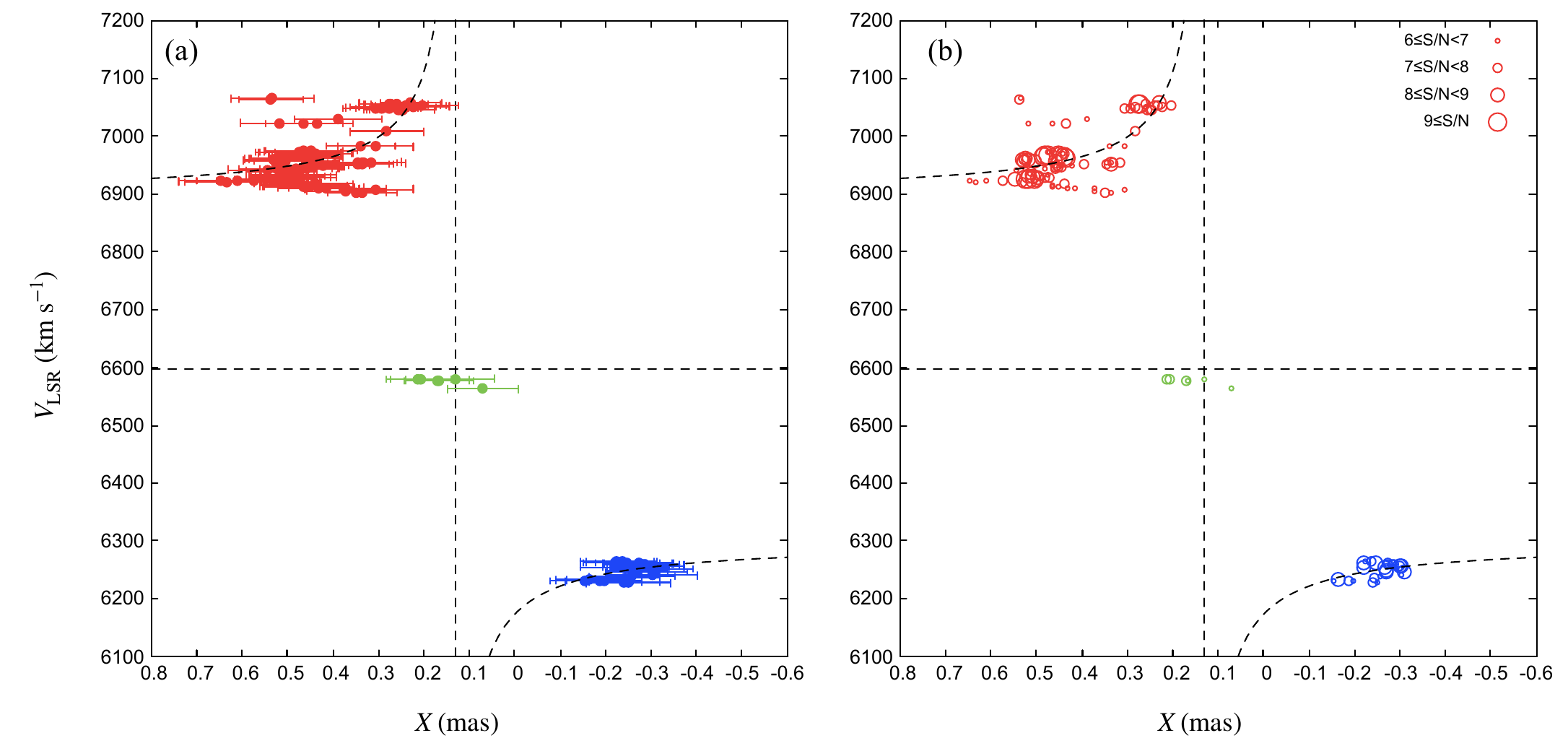}
  \end{center}
  \caption{\textrm{
  (a) Position-velocity diagram along the $X$ axis
  {in figure \ref{fig:pa}}.
  (b) Same as (a) but without error bars.
  {The size of the circles represents the signal-to-noise ratio.}
  Vertical dotted lines indicate the dynamical center of the maser disk ($X_0={0.13}\>\mathrm{mas}={0.06}\>\mathrm{pc}$) and
  horizontal dotted lines the system velocity ($V_\mathrm{sys}={6598}$\km).
  Dotted curves are the result of least-squares fitting of the high-velocity maser spots to equation (\ref{eq:vrot}) in the text.
  }
  }
  \label{fig:pv}
\end{figure*}

Figure \ref{fig:pa} shows the distribution of maser spots in the frame of the coordinates of $X$ along $\mathrm{PA}={\timeform{8.6D}}$.
Figure \ref{fig:pv} shows the position-velocity diagram along the $X$ axis.
To investigate the rotation of the maser disk,
we fitted the power law,
  \begin{equation}
    \label{eq:fit}
    V_\mathrm{LSR}=A(X-X_0)^{\alpha}+V_\mathrm{sys},
  \end{equation}
to the high-velocity features in the diagram, using the least squares method,
obtaining
$A={303} \, \pm \,{12} \, \mathrm{km\>s^{-1}}$,
$X_0 = {0.18} \, \pm \, {0.03} \,\mathrm{mas}$,
$\alpha = {-0.14} \, \pm \, {0.04}$, and
$V_\mathrm{sys} = {6592} \, \pm \, {4} \> \mathrm{km\>s^{-1}}$.

The power-law index $|\alpha|={0.14} \, \pm \, {0.04}$ is smaller than that of the Keplerian rotation ($V \propto r^\alpha$ with $\alpha=-0.5$){, indicating} that compared to the Keplerian, the rotation velocity of the maser disk decreases more slowly to the outer side.
Non-Keplerian rotation suggests that the mass does not reduce to a central point source;
instead, it extends outward across the masing region.


\begin{figure}[h]
\begin{center}
  \includegraphics[width=80mm]{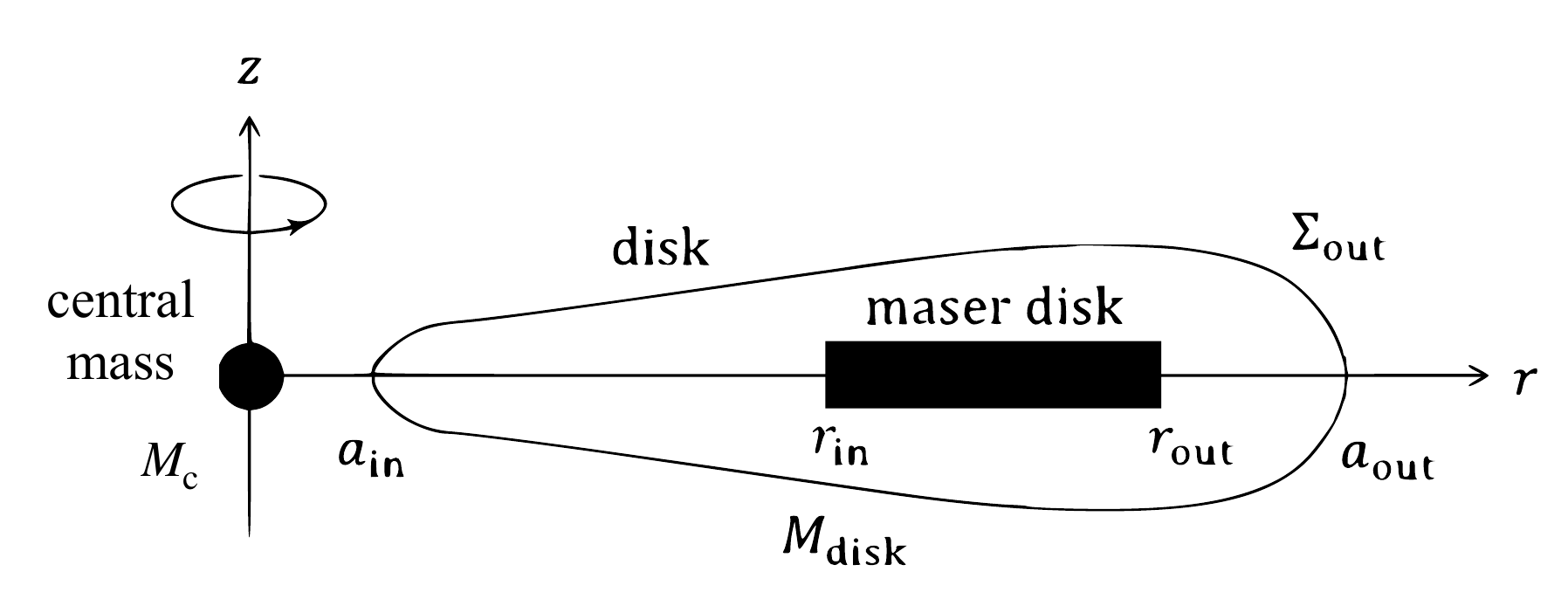}
\end{center}
  \caption{
  \textrm{{
  Schematic view of a two components mass model.
An axisymmetrical gaseous disk including the maser disk rotates around a central point mass.}
}
}
  \label{fig:bh}
\end{figure}
\begin{figure*}
  \begin{center}
  \includegraphics[width=160mm]{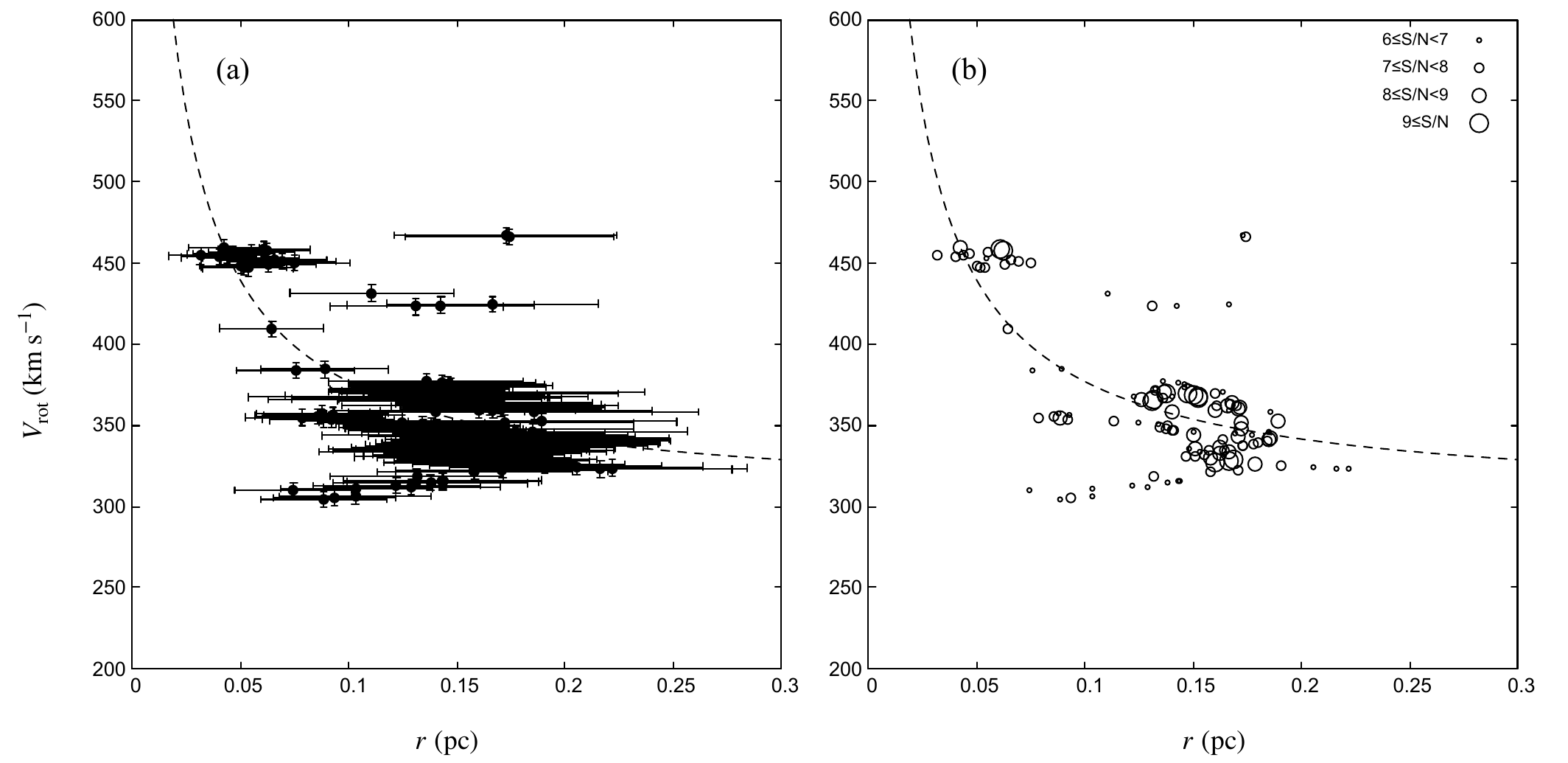}
  \end{center}
  \caption{\textrm{
  {
  (a)
  A rotation curve of the {maser} disk obtained from figure \ref{fig:pv}.
  The {abscissa represents} the distance from the dynamical center of the maser disk,
  $r \equiv |X-X_0|$,
  and the {ordinate} the rotation velocity,
  $V_\mathrm{rot}=|V_\mathrm{LSR}-V_\mathrm{sys}|$ where $X_0={0.13}\>\mathrm{mas}$ $({0.06}\>\mathrm{pc})$ and $V_\mathrm{sys}={6598}$\km.
  The dotted line indicates the fitting result using equation (\ref{eq:vrot}),
  same as that in figure \ref{fig:pv}.
  (b) Same as (a) but without error bars.
  {The size of the circles indicates the signal-to-noise ratio.}
  }
}
  }
  \label{fig:rot}
\end{figure*}


\subsection{Mass distribution}\label{sub:mass}
\subsubsection{{Two-component mass model}}
{
While the extended mass distribution may have spherical symmetry such as a high-density star cluster (e.g., \cite{Herrnstein2005}; \cite{Kuo2011}),
{such a} star cluster cannot be visualized in the presence of a dense gas disk.
Following \citet{Hure2002} and \citet{Hure2007}, we consider an axisymmetric mass distribution model composed of a central point source
such as a black hole and a disk that contains the maser disk orbiting the point source.
We do not correct for gravitational red-shift and doppler shift due to the special relativity,
as these corrections are extremely small in this galaxy ($\Delta V<1$\km).}

{Figure \ref{fig:bh} illustrates the two-component mass model, in which
an axisymmetrical gaseous disk with inner edge $a_\mathrm{in}$ and outer edge $a_\mathrm{out}$,
including the maser disk at $r_\mathrm{in}<r<r_\mathrm{out}$,
rotates around a central point mass of \mc.
Assuming centrifugal balance, the rotational velocity,
$V_\mathrm{rot}$,
of the gas located at a cylindrical distance $r$ from the center is defined (ignoring the pressure gradient) as}
  \begin{equation}
    \frac{{V_\mathrm{rot}}^2}{r}=\frac{GM_\mathrm{c}}{r^2}+\frac{d\psi}{dr},
  \end{equation}
where $\psi$ is the gravitational potential of the disk,
and $G$ the gravitational constant.
In case of a flat and thin disk,
and
$a_\mathrm{in}\ll{} a_\mathrm{out}$, $r_\mathrm{out}\ll{} a_\mathrm{out}$ and $\beta=-1$
of equation (\ref{eq5})
(see the Appendix), i{.}e., the Mestel disk,

\begin{equation}
  V_\mathrm{rot}(r)=\sqrt{\frac{GM_\mathrm{c}}{r}+2\pi{}G\Sigma_\mathrm{out}a_\mathrm{out}},
  \label{eq:vrot}
\end{equation}
{where $V_\mathrm{rot}=|V_\mathrm{LSR}-V_\mathrm{sys}|$, $r=|X-X_0|$,
and $\Sigma_\mathrm{out}$ the surface density at $a_\mathrm{out}$.}

{
We performed the least squares fitting for the maser points in figure \ref{fig:pv},
using equation (\ref{eq:vrot}) with four parameters of $V_\mathrm{sys}$, $X_0$, \mc{} and
$\Sigma_\mathrm{out}a_\mathrm{out}$ as the dotted line in figure \ref{fig:pv}{, where we assume that the disk is edge-on and high-velocity masers in the maser disk lie on the mid-line of the disk perpendicular to the line of sight}.
Using the resultant values of
{the fitting,}
$V_\mathrm{sys}={6598}\,\pm\,{5}$\km{} and $X_0={0.13}\,\pm\,{0.03}\>\mathrm{mas}$.
{
A rotation curve of
}
figure \ref{fig:rot} was drawn from figure \ref{fig:pv}}{, and $r_\mathrm{in}={0.031}$\>pc, $r_\mathrm{out}={0.222}\>$pc, $V_\mathrm{in}={454}$\km{} at \rin, and $V_\mathrm{out}={324}$\km{} at \rout{}.}
{The} disk mass at the distance between $r_1$ and $r_2$ from the center of rotation can be obtained as
\begin{equation}
  M_\mathrm{disk}=\int_{r_1}^{r_2}\Sigma(r)2\pi{}rdr
  =2\pi{}\Sigma_\mathrm{out}a_\mathrm{out}(r_2-r_1),
  \label{eq:mdisk}
\end{equation}
using equation (\ref{eq5})
{with $\beta=-1$}.
As results of fitting,
the masses of {$M_\mathrm{c}$},
the disk within $r_\mathrm{out}$, $M_{\mathrm{disk}}{(\leq r_\mathrm{out})}$,
and the maser disk between $r_\mathrm{in}$ and $r_\mathrm{out}$,
$M_\mathrm{disk}{\mathrm{(maser)}}$ are
\begin{eqnarray}\label{eq:mass}
  M_\mathrm{c}=({1.2}\,\pm\,{0.4})\times10^6 \Mo, \nonumber\\
  M_{\mathrm{disk}}{(\leq r_\mathrm{out})}=({4.7}\,\pm\,{1.5})\times10^6 \Mo, \\
  M_\mathrm{disk}{\mathrm{(maser)}}=({4.0}\,\pm\,{1.5})\times{10^6} \Mo,\nonumber
\end{eqnarray}
respectively{, where the errors include the uncertainty of the galaxy distance ($88.9\,\pm\,3.6\>\mathrm{Mpc}$)}.
{The thickness of the maser disk {is} $2H{=0.035\,\pm\,0.07}$\>pc ($H/r_\mathrm{in}{=0.56}$) but this is the upper limit, because the synthesized beam with {$\mathrm{PA}=\timeform{10D}$} ({figures} \ref{fig:spots} {and \ref{fig:pa}}) spread the apparent distribution of the maser spots.
}

\subsubsection{{Possible systemic errors}}
We assumed an edge-on maser disk with an inclination angle of $i=\timeform{90D}$ to fit equations (\ref{eq:fit}) and (\ref{eq:vrot}) to the maser distribution in the position-velocity diagram in figure \ref{fig:pv}.
If the disk is not edge-on but tilted with a constant $i$ ($<\timeform{90D}$) throughout the disk, then \vlsr{} in equation (\ref{eq:fit}) and \vrot{} in equation (\ref{eq:vrot}) must be divided by $\sin(i)$.
The maser disk of this galaxy may be tilted, because the mean position of the systemic velocity maser features in figures \ref{fig:spots} and \ref{fig:pa} does not appear on the dotted line connecting the red- and blue-shifted features but off to the west or $Y>0$.
However, determining the inclination angle $i$ is challenging, because the maser spots of the systemic features spread, and their distances from the disk center are unknown.
In addition, even if the maser disk inclines, the correction by $\sin(i)$ would be small; most megamaser disks of other galaxies{, except NGC\,1068 ($i\approx\timeform{75D}$;  \cite{Gallimore2023}),} have shown $i>\timeform{80D}$ (e.g., \cite{Miyoshi1995}; \cite{Herrnstein1996}; \cite{Kuo2011}; \cite{Pesce2020}) and $\sin(i>\timeform{80D})>0.985$ which is smaller than $\sin(i=\timeform{90D})$
by only $<1.5\%$.
The power $\alpha$ in equation (\ref{eq:fit}) does not change.
The rotational velocity \vrot{} in equation (\ref{eq:vrot}) may increase by $<1.5\%$; hence, \mc{} and \mdisk{} increase by $<(1.5\%)^2=2.3\%$, whereas the ratio of $M_\mathrm{disk}/M_\mathrm{c}$ does not change.

If the maser disk warps, i.e., the inclination angle varies with the distance from the disk center,
then the power $\alpha$ and $M_\mathrm{disk}/M_\mathrm{c}$ ratio could be affected;
however correction of inclination is small for $i>\timeform{80D}$.
\citet{Herrnstein2005} analyzed the effects of inclination and warping of the maser disk on its rotation curve, and hence, the black hole and disk masses in NGC\,4258 in detail.
{In \ngc, the distribution of the red-shifted features in figure \ref{fig:spots} (and figure \ref{fig:pa}) is tilted toward $\mathrm{PA}\sim\timeform{-10D}$.
This may indicate a warp of the maser disk, but the effect is indistinguishable from the effect of the tilted and elongated beam, because the PA is close to that of the synthesized beam.
The systemic velocity features also distribute toward the same PA. Further VLBI observations with higher sensitivity and angular resolution are required to examine a warp of the disk.}

If the high-velocity maser features do not lie on the disk's mid-line (intersection between disk plane and sky plane),
then the observed velocities projected onto the line-of-sight (LOS) are smaller than the true orbital velocities.
\citet{Kuo2011}, \citet{Reid2013}, \citet{Humphreys2013}, and \citet{Gao2016} showed that high-velocity features can lie in regions deviating from the disk's mid-line by approximately $\timeform{10D}\mbox{--}\timeform{20D}$ in several maser disks.
Deviations of maser positions in the outer region from the disk's mid-line cause faster decreasing of the LOS rotation speed outward than the true rotation curve, resulting in a smaller power index $\alpha$ (larger $|\alpha|$) and hence a smaller $M_\mathrm{disk}/M_\mathrm{c}$ ratio.
Deviations in the inner region induce an apparent larger $\alpha$ (smaller $|\alpha|$) and a larger $M_\mathrm{disk}/M_\mathrm{c}$ ratio.
\citet{Kuo2018} estimated black hole and disk masses in six galaxies with maser disks showing nearly perfect Keplerian rotation,
both without and with corrections for deviations maser position from disk mid-lines using equation (12) in \citet{Hure2011} [equation (\ref{eq19}) in this paper].
The corrected case yielded black hole masses $+5.7\%\mbox{--}-8.1\%$
larger and disk masses several to ten times smaller than the uncorrected case,
with better reduced $\chi^2$.

{
In NGC\,1068, \citet{Greenhill1996}, \citet{Lodato2003}, and \citet{Morishima2023} reported a sub-Keplerian maser disk with $|\alpha|=0.24\mbox{--}0.35$ using the rotation curve, and \citet{Lodato2003} and \citet{Hure2011} showed a massive disk with $M_\mathrm{disk}/M_\mathrm{BH}\sim1\mbox{--}1.5$
by analyzing the rotation curve.
However, \citet{Gallimore2023} obtained Keplerian rotation and a low mass disk with $M_\mathrm{disk}/M_\mathrm{BH}=6.5\times10^{-3}$, where the disk mass, $M_\mathrm{disk}$, was estimated on the basis of disk stability arguments, adopting a spiral arm model of maser distribution.
In most cases of water maser galaxies, maser spectra exhibit the triple-peaked structure corresponding to red-shifted, blue-shifted, and systemic velocity features, and the distribution of the three maser groups separates in their distribution on the sky and in the position velocity diagram.
The high velocity features red- and blue-shifted along the major axis of the maser disk also constitute terminal velocities of the disk rotation indicating a rotation curve.
In case of NGC\,1068, however, the systemic velocity features and the red-shifted features are continuously connected in the P-V diagram as well as the spatial distribution on the sky, resulting in difficulty in  distinguishing the red-shifted features from the systemic features.
In addition, maser spots in the spiral arms proposed by \citet{Gallimore2023} may not trace terminal velocities of the rotating disk, lacking some maser components along the major axis.
Such effects may induce the apparent non-Keplerian rotation and the massive disk.
}

For \ngc, maser disk rotation with $|\alpha|=0.14\,\pm\,0.04$ in equation (\ref{eq:fit}) is far from Keplerian ($|\alpha|=0.5$), indicating significant disk mass.
However{, the small disk size of $r=0.031\mbox{--}0.222\>\mathrm{pc}$ at the far distance of $D_\mathrm{A}=88.9\>\mathrm{Mpc}$ makes its apparent size very small ($r=0.07\mbox{--}0.52\>\mathrm{mas}$).
In addition, the flux densities of individual maser spots are weak (table \ref{tab:posi}).
These effects increase the errors in maser positions in the position-velocity diagram (figure \ref{fig:pv}) and the rotation curve (figure \ref{fig:rot}).
To determine more accurate rotation of the maser disk and masses of the black hole and disk,} further VLBI observations with higher sensitivity and angular resolution {are needed}.

\subsection{{Black hole}}\label{sub:bh}
{
To examine whether the central mass,  \mc{} is a black hole, we calculated the central mass density, following the method demonstrated by \citet{Ishihara2001}.
The mass inside the inner radius, \rin{} is given by $M = V_\mathrm{in}^2r_\mathrm{in}/G = ({1.51}\,\pm\,{0.03})\times10^6\Mo$,
assuming a spherical distribution of matter.
The mean volume density inside the inner radius can be calculated to be
$\rho=3M/(4\pi{}r_\mathrm{in}^3)=({1.2}\,\pm\,{0.5})\times{10^{10}}\Mo\>\mathrm{pc^{-3}}$ which is comparable with those of NGC\,4258 (e.g., \cite{Miyoshi1995}; \cite{Moran1995}),
NGC\,2273, UGC\,3789, NGC\,2960, NGC\,4388, and NGC\,6323 (\cite{Kuo2011}), and IC\,2560 (\cite{Yamauchi2012}),
strongly suggesting that the central mass is a {supermassive} black {hole} [see more detailed discussion in, e.g., \citet{Maoz1995}; \citet{Ishihara2001}; \citet{Kuo2011}].

The {black hole mass} of {$M_\mathrm{BH}(=M_\mathrm{c})={1.2}\times10^6\Mo$ of \ngc} is the same as those of NGC\,3079
{
[$\sim10^6\Mo$, \cite{Trotter1998}; ($2\mbox{--}3)\times10^6\Mo$, \cite{Yamauchi2004}; $2\times10^6\Mo$, \cite{Kondratko2005}],
NGC\,4945 ($1.4\times10^6\Mo$, \cite{Greenhill1997}),
IC\,2560 ($2\times10^6\Mo$, \cite{Yamauchi2012}),
{Circinus} galaxy [$(1.7\,\pm\,0.3)\times10^6\Mo$, \cite{Greenhill2003}],
and J0437+2456 [$(2.9\,\pm\,0.3)\times10^6\Mo$, \cite{Gao2017}],}
though it is one order of magnitude smaller than those of NGC\,4258 et al.

The Schwarzschild radius for the black hole mass is
$r_\mathrm{s}=2GM_\mathrm{BH}/c^2={1.1}\times10^{-7}\>\mathrm{pc}$,
where $c$ is the speed of light,
and thus  $r_\mathrm{in}={2.8}\times10^5\,r_\mathrm{s}$ and $r_\mathrm{out}={2.0}\times10^6\,r_\mathrm{s}$.
{

\subsection{{Disk stability}}
{
The fitting results of equation (\ref{eq:mass}) shows that the maser disk is massive with $M_\mathrm{disk}(\leq r_\mathrm{out})/M_\mathrm{BH}=3.9\,\pm\,1.8$.
We examine the gravitational stability of the disk by evaluating the Toomre $Q$ parameter, $Q \equiv \kappa c_\mathrm{s}/\pi G \Sigma$, where $\kappa$ is the epicyclic frequency, $c_\mathrm{s}$ the sound speed, $G$ the gravitational constant, and $\Sigma$ the surface density of the disk.
Our forthcoming paper (Deepshikha et al.) calculates the $Q$ parameter of 23 maser disks including \ngc.
The $Q$ value of \ngc{} is $Q=0.7\,\pm\,1.3$ and $0.5\,\pm\,0.6$ at $r_\mathrm{in}$ and $r_\mathrm{out}$, respectively, using $\kappa=(2.9\,\pm\,2.9)\times10^{-10}\>\mathrm{s^{-1}}$ at $r_\mathrm{in}$ and $(0.40\,\pm\,0.14)\times10^{-10}\>\mathrm{s^{-1}}$ at $r_\mathrm{out}$,
$c_\mathrm{s}$ the sound speed but adopted from the velocity dispersion ($\sigma_\mathrm{V}=30\,\pm\,24$\km),
$G=0.00430091(\>\mathrm{km\>s^{-1}})^2\Mo\>\mathrm{pc}$ in the unit, and $\Sigma=(2.6\,\pm\,3.2)\times10^{7}\Mo\>\mathrm{pc^{-2}}$ at $r_\mathrm{in}$ and $(0.55\,\pm\,0.43)\times10^{7}\Mo\>\mathrm{pc^{-2}}$ at $r_\mathrm{out}$.
The values of $Q<1$ indicate that the maser disk is gravitationally unstable, warranting a massive disk, although its error is large and hence more accurate observation is desired.
}

{
The forthcoming paper also shows that the maser disk of IC\,1481 is non-Keplerian ($\alpha=-0.15\,\pm\,0.07$), has large disk mass ($M_\mathrm{disk}/M_\mathrm{BH}=2.7\,\pm\,1.0$), and is unstable ($Q=0.14\,\pm\,0.13$ at $r_\mathrm{in}$ and $0.12\,\pm\,0.10$ at $r_\mathrm{out}$).
The position-velocity diagram or rotation curve of the maser disk in \citet{Mamyoda2009} shows a cluster of maser spots with a large velocity dispersion of $\Delta V=20$\km{} ($6106<V_\mathrm{rad, LSR}<6126$\km) at $r\approx10\>\mathrm{pc}$ from the dynamical center.
The maser clustering in the region of the wide velocity range suggests outflow or expansion of dense gas in the maser disk, which is inducing strong red-shifted maser emission probably caused by shock (Mamyoda et al., private communication).
}

{
The maser disk of NGC\,3079 shows sub-Keplerian ($\alpha=-0.36\,\pm\,0.29$) and large disk mass ($M_\mathrm{disk}/M_\mathrm{BH}=0.24\,\pm\,1.09$), although the errors are large (Deepshikha et al.).
The maser disk is also unstable ($Q=0.14\,\pm\,0.13$ at $r_\mathrm{in}$ and $0.12\,\pm\,0.10$ at $r_\mathrm{out}$).
\citet{Yamauchi2004} showed two peculiar local maser groups with large velocity ranges of $\Delta V=74$\km{} ($V_\mathrm{rad, LSR}=937\mbox{--}1011$\km) at $r\approx0.6\>\mathrm{pc}$ and $\Delta V=32$\km{} ($V_\mathrm{rad, LSR}=1002\mbox{--}1034$\km) at $r\approx1.2\>\mathrm{pc}$.
The latter maser group suggests an expanding shell with expanding energy of $E\sim10^{45}(n_\mathrm{H}/10^8\>\mathrm{cm}^{-3})(v_\mathrm{exp}/10\>\mathrm{km\>s^{-1}})^2\>\mathrm{erg}$, where $n_\mathrm{H}$ is the hydrogen density of the shell and $v_\mathrm{exp}$ the expanding velocity.
Since the expanding energy is less than the typical energy of a supernova explosion ($E\sim10^{51}\>\mathrm{erg}$), the local expansion of the dense gas may be caused by stellar wind of massive stars.
The two local wide velocity groups show exceptionally intense maser emission at the blue-shifted velocities, caused by strong shock.
}

{
In IC\,1481 and NGC\,3079, massive stars may be forming in the unstable maser disks. In \ngc, such outflow/expansion is not seen in the maser disk, but the disk shows non-Keplerian rotation ($\alpha=-0.14$) and a unstable disk like IC\,1481 and NGC\,3079.
The $Q$ values of Keplerian maser disks ($\alpha\approx-0.5$) in other galaxies such as NGC\,4258 are larger than one (Deepshikha et al.), indicating stable disks. The forthcoming paper will analyze the gravitational stability of maser disks in more detail.
}

\subsection{X-ray emission from nucleus and mass-accretion rates}\label{sub:xray}

\citet{Terashima2015} performed a spectral analysis of the data obtained from the XMM-Newton Science Archive,
and obtained the 2--10\>keV luminosity of
$L$(2--10\>keV)=(3.7--4.9)$\times10^{40}\>\mathrm{erg\>s^{-1}}$,
correcting absorption.

The Eddington luminosity of \ngc{} can be calculated using the central mass of the galaxy
({1.19}$\times10^6 \Mo$; see subsection \ref{sub:mass}), as
\begin{eqnarray}
  L_\mathrm{E}& =&4\pi{}GM_\mathrm{BH}m_\mathrm{H}c/\sigma_\mathrm{T} \nonumber\\
  &=& {1.5}\times10^{44}\>\mathrm{erg\>s^{-1}}
\end{eqnarray}
where $m_\mathrm{H}$ is the mass of hydrogen atom
and $\sigma_\mathrm{T}$ the Thomson cross section
($\sigma_\mathrm{T}=6.65\times10^{-25}\>\mathrm{cm^2}$).
Thus, the normalized luminosity becomes
$L$(2--10\>keV)/$L_\mathrm{E}=({2.4}\mbox{--}{3.2})\times10^{-4}$.

From the X-ray luminosity,
we can estimate the mass-accretion rate of the central black hole (e.g., \cite{Ishihara2001}).
We assumed a standard accretion model around a Schwarzschild black hole  to estimate the mass-accretion rate of \ngc.
In the model,
a stable disk is formed up to 3 times the Schwarzschild radius, $r_\mathrm{s}(=2GM_\mathrm{BH}/c^2\approx{3.5}\times10^6\>\mathrm{km})$.
At a radius of $<3r_\mathrm{s}$,
the mass cannot form a stable circular orbit and falls onto the central black hole.
Half of the energy emitted until then is considered to be radiation, and since
\begin{eqnarray}\label{eq:7}
  L &\simeq& \frac{1}{2}\frac{d}{dt}(\frac{GM_\mathrm{BH}M}{3r_\mathrm{s}})\nonumber\\
  &=&\frac{1}{2}G \frac{M_\mathrm{BH}\dot{M}}{3r_\mathrm{s}} \nonumber\\
  &=& \frac{1}{12}\dot{M}c^2,
\end{eqnarray}
the mass-accretion rate, $\dot{M}$ for this galaxy, becomes
\begin{eqnarray}
  \dot{M} &\simeq& 12\frac{L}{c^2} \nonumber\\
  &=& (0.8\mbox{--}1.0)\times10^{-5}\>M_\odot\mathrm{yr^{-1}},
\end{eqnarray}
for $L=L$(2--10\>keV).
Compared to other megamaser galaxies
(\cite{Ishihara2001}),
the mass-accretion rate of \ngc{} is comparable to that of NGC\,4258,
and is several orders of magnitude smaller than those of NGC\,1068, NGC\,4945 and the Circinus galaxy.

The low Eddington ration [$L/L_\mathrm{E}=({2.4}\mbox{--}{3.2})\times10^{-4}$] might indicate that the inner part of the accretion disk becomes radiatively in efficient accretion flow.
In such a case, the radiation efficiency would be lower than that estimated above, and the $\dot{M}$ value might be underestimated.


\section{Summary}
Water maser emission from \ngc was observed using the VLBI.
The results are summarized below:
\begin{enumerate}
  \item In addition to the known red- and blue-shifted features,
  systemic velocity features are clearly detected.
  The relative velocities from the systemic velocity are ${305}\mbox{--}{468}$ and ${333}\mbox{--}{370}$\km{} for the red- and blue-shifted features, respectively.

  \item The maser features are emitted from a nearly edge-on disk with a position angle of $\mathrm{PA}={\timeform{8.6D}}\,\pm\,{\timeform{0.8D}}$.
  The radius and thickness of the maser disk are {0.031}--0.222\>pc and $2H<0.035$\>pc, respectively.
  The rotation velocity is $V_\mathrm{rot}={324}\mbox{--}{454}$\km.
  The maser disk is almost perpendicular to the kpc-scale galactic disk.

  \item The rotation curve of the maser disk falls more gradually ($V_\mathrm{rot} \propto r^{{-0.14} \, \pm \, {0.04}}$) than the Keplerian rotation, indicating a self-gravitating disk.
  Based on a model (in which the mass is composed of a central black hole and surrounding disk including the maser disk), the determined central black hole mass,
  mass inside the outer radius of the disk,
  and mass of the maser disk are
  $M_\mathrm{BH}=({1.2}\,\pm\,{0.4})\times10^6 \Mo$,
  $M_{\mathrm{disk}}{(\leq r_\mathrm{out})}=({4.7}\,\pm\,{1.5})\times{10^{6}}\Mo$, and
  $M_\mathrm{disk}{\mathrm{(maser)}}=({4.0}\,\pm\,{1.5})\times{10^{6}}\Mo$, respectively.
  The mean volume density of the center is $\rho\approx{10^{10}}\Mo\>\mathrm{pc^{-3}}$,
  suggesting the existence of a supermassive black hole.

  \item  The mass-accretion rate for the central black hole, calculated from the  2--10\>keV luminosity using the standard accretion model, is
  $(0.8\mbox{--}1.0)\times10^{-5}\>M_\odot\mathrm{yr^{-1}}$.
  The Eddington ratio is $({2.4}\mbox{--}{3.2})\times10^{-4}$.

\end{enumerate}

\bigskip
\begin{ack}
  The authors thank Dr. K. Mamyoda and Mr. K. Isami whose calculations were helpful for deriving equation
  {(\ref{eq:vrot}).
  We also thank Ms. Deepshikha for calculation of $Q$ values}.
  This research used the NASA/IPAC Extragalactic Database (NED),
  which is funded by the National Aeronautios and operated by the California Institute of Technology.
\end{ack}

\section*{Supplementary data}
Supplementary data are available in the online version of this article.
E-table 1.

\appendix
\section*{Derivation of equation {\ref{eq:vrot}}}

We considered {the two-components} mass model of a central black hole with mass \mbh{} and a gaseous disk including the maser disk, as shown in figure \ref{fig:bh},
where the radii of inner and outer edges of the gaseous disk are \ain{} and \aout{,} and the maser disk is located at $r_\mathrm{in}<r<r_\mathrm{out}$ ($  a_\mathrm{in}<r_\mathrm{in}$, $r_\mathrm{out}<a_\mathrm{out}$).
The gaseous disk was assumed axially symmetrical, flat, pressureless, and steady.

At centrifugal {equilibrium},
the rotation speed $V_\mathrm{rot}$ of the gas at the distance from the center $r$ in the disk is given by
  \begin{equation}
    \frac{{V_\mathrm{rot}}^2}{r}=\frac{GM_\mathrm{BH}}{r^2}+\frac{d\psi_\mathrm{disk}}{dr}, \label{eq1}
  \end{equation}
where $G$ is the gravitational constant and $\psi_\mathrm{disk}$ the gravitational potential of the disk.
The equation represents the theoretical rotation velocity $V_\mathrm{rot}(r)$ formed by both the black hole and the gaseous disk and in principle, yields \mbh{} and the disk mass \mdisk{} for \rin, \rout{} and mass distribution of the disk as soon as two data points $(r, V_\mathrm{rot})$ are known,
but the inversion is not analytical.
Under certain circumstances however, a simple and tractable formula for the function $\psi_\mathrm{{disk}}$ can be deduced (\cite{Hure2007}; \cite{Hure2008}) so that inversion is straightforward.

We followed the method by \citet{Hure2007}, \citet{Hure2008}, \citet{Mamyoda2010}, {and} \citet{Hure2011}.
In the case of a flat and thin axisymmetric disk, the potential $\psi(r)$ due to the disk at a distance $r$ from the center in the plane is (e.g., \cite{Durand1953})
  \begin{equation}
    \psi(r)= -2G \int_{a_\mathrm{in}}^{a_\mathrm{out}} \sqrt{\frac{a}{r}} \Sigma(a) m \bm{K}(m) da
    \label{eq2},
  \end{equation}
where $\Sigma(a)$ is the surface density of the disk at the distance {of} $a$ and $\bm{K}(m)$ is the complete elliptic integral of the first kind,
  \begin{equation}
    \bm{K}(m)=\int_{0}^{\frac{\pi}{2}} \frac{1}{\sqrt{1-m^2\sin^2{\phi}}}d\phi,
    \label{eq3}
  \end{equation}
and $m$ is the modulus with
  \begin{equation}
    m = \frac{2 \sqrt{ar}}{a+r} \qquad (0 \leq m \leq 1).
    \label{eq4}
  \end{equation}
We consider the surface density distribution in the disk as
  \begin{equation}
    \Sigma(a)=\Sigma_\mathrm{out}\left(\frac{a}{a_\mathrm{out}}\right)^\beta,
    \label{eq5}
  \end{equation}
where $\Sigma_\mathrm{out}$ is the surface density at \aout{} and $\beta$ a real exponent.
The modulus of the complete elliptic integral can change as (\cite{Gradshteyn1965})
  \begin{equation}
    \bm{K}\left(\frac{2\sqrt{x}}{1+x}\right)=(1+x)\bm{K}(x)
    \qquad (x\leq1).
    \label{eq6}
  \end{equation}
We divide the integral interval of equation  (\ref{eq2}) into $a_\mathrm{in}-r$ and $r-a_\mathrm{out}\;(a_\mathrm{in}<r<a_\mathrm{out})$ as
  \begin{eqnarray}
    \int_{a_\mathrm{in}}^{a_\mathrm{out}} \sqrt{\frac{a}{r}} \Sigma(a) m \bm{K}(m) da
    =\int_{a_\mathrm{in}}^{r} \sqrt{\frac{a}{r}} \Sigma(a) m \bm{K}(m) da \nonumber \\
    + \int_{r}^{a_\mathrm{out}} \sqrt{\frac{a}{r}} \Sigma(a) m \bm{K}(m) da.
    \label{eq7}
  \end{eqnarray}
When we define $v\equiv r/a \;(r \leq a \leq a_\mathrm{out}; v\leq 1)$ and $v_\mathrm{out}\equiv r/a_\mathrm{out}$ ,
the second term of the right side of equation (\ref{eq7}) is
  \begin{equation}
    \int_{r}^{a_\mathrm{out}} \sqrt{\frac{a}{r}} \Sigma(a) m \bm{K}(m) da
    =-2r\Sigma_\mathrm{out}v_\mathrm{out}^{\beta}\int_{1}^{v_\mathrm{out}}\frac{\bm{K}(v)}{v^{2+\beta}}dv,
    \label{eq8}
  \end{equation}
using equations (\ref{eq4}), $m=2\sqrt{v}/(1+v)$, (\ref{eq5}),
$\Sigma(a)=\Sigma_\mathrm{out}v_\mathrm{out}^{\beta}/v^{\beta}$, and (\ref{eq6}) $(x=v)$.
When we define $u\equiv a/r \; (a_\mathrm{in} \leq a \leq r; u\leq 1)$ and $u_\mathrm{in}\equiv a_\mathrm{in}/r$,
the first term of the right side of equation (\ref{eq7}) is
  \begin{equation}
    \int_{a_\mathrm{in}}^{r} \sqrt{\frac{a}{r}} \Sigma(a) m \bm{K}(m) da
    =2r\Sigma_\mathrm{out} v_\mathrm{out}^{\beta}\int_{u_\mathrm{in}}^{1}\bm{K}(u){u^{1+\beta}}du,
    \label{eq9}
  \end{equation}
using equations (\ref{eq4}), $m=2\sqrt{u}/(1+u)$, (\ref{eq5}),
$\Sigma(a)=\Sigma_\mathrm{out}v_\mathrm{out}^{\beta}/u^{\beta}$, and (\ref{eq6}) $(x=u)$.
From equations (\ref{eq7}), (\ref{eq8}), and (\ref{eq9}), equation (\ref{eq2}) becomes
\begin{eqnarray}
  \psi_\mathrm{disk}(r)&=&
  -4Gr\Sigma_\mathrm{out}v_\mathrm{out}^\beta \nonumber \\
  &\times& \left\{\int_{u_\mathrm{in}}^1 \bm{K}(u)u^{1+\beta}du
  -\int_1^{v_\mathrm{out}} \frac{\bm{K}(v)}{v^{2+\beta}}dv
  \right\}.
  \label{eq10}
\end{eqnarray}

Differential of equation (\ref{eq10}) is given by
  \begin{eqnarray}
    \frac{d\psi_\mathrm{disk}}{dr} &=&(1+\beta)\frac{\psi_\mathrm{disk}}{r} \nonumber \\
    &-& 4Gr\Sigma_\mathrm{out}v_\mathrm{out}^\beta  \nonumber \\
    &\times&
    \frac{d}{dr}
    \left\{\int_{u_\mathrm{in}}^1\bm{K}(u)u^{1+\beta}du- \int_1^{v_\mathrm{out}}
    \frac{\bm{K}(v)}{v^{2+\beta}}dv\right\}.
    \label{eq11}
  \end{eqnarray}
If the equation,
  \begin{equation}
    \frac{d}{dx}\int_{p(x)}^{q(x)}f(t)dt=f(q(x))\frac{dq}{dx}-f(p(x))\frac{dp}{dx}, \label{eq12}
  \end{equation}
is applied to equation (\ref{eq11}),
  \begin{eqnarray}
    \frac{d}{dr}\int_{u_\mathrm{in}}^1 \bm{K}(u)u^{1+\beta}du
    &=&-\bm{K}(u_\mathrm{in})u_\mathrm{in}^{1+\beta}\frac{du_\mathrm{in}}{dr}\nonumber \\
    &=&\frac{1}{r}\bm{K}(u_\mathrm{in})u_\mathrm{in}^{2+\beta},
    \label{eq13}
  \end{eqnarray}
  \begin{eqnarray}
    \frac{d}{dr}\int_{1}^{v_\mathrm{out}} \frac{\bm{K}(v)}{v^{2+\beta}}dv
    &=&\frac{\bm{K}(v_\mathrm{out})}{v_\mathrm{out}^{2+\beta}} \frac{dv_\mathrm{out}}{dr} \nonumber \\
    &=&\frac{1}{r}\frac{\bm{K}(v_\mathrm{out})}{v_\mathrm{out}^{1+\beta}},
    \label{eq14}
  \end{eqnarray}
so that
 \begin{eqnarray}
   \frac{d\psi_\mathrm{disk}}{dr}
   &=&(1+\beta)\frac{\psi_\mathrm{disk}}{r} \nonumber \\
   &-&4G\Sigma_\mathrm{out}(\frac{a_\mathrm{in}}{a_\mathrm{out}})^\beta \bm{K}(u_\mathrm{in})u_\mathrm{in}^2
   +4G\Sigma_\mathrm{out}\frac{\bm{K}(v_\mathrm{out})}{v_\mathrm{out}}.
   \label{eq15}
 \end{eqnarray}
Here we assume (1) $a_\mathrm{in}\ll a_\mathrm{out}$, (2) $r_\mathrm{out}\ll a_\mathrm{out}$, and (3) the Mestel disk {\citep{Mestel1963}}, i.e., $\beta=-1$.
Since the Schwarzschild radius is $r_\mathrm{s}\sim10^{-7}$\>pc (section \ref{sub:xray}),
$a_\mathrm{in}\sim3r_\mathrm{s}\sim10^{-7}\>\mathrm{pc}\ll r_\mathrm{out}=0.22\>\mathrm{pc}<a_\mathrm{out}$,
satisfying assumption (1).
If $r_\mathrm{out}\sim a_\mathrm{out}$,
$v_\mathrm{out}\sim1$.
In this case, $\bm{K}(v_\mathrm{out})$ is extremely large (almost infinite; figure \ref{fig1}),
so that the gravity $d\psi_\mathrm{disk}/dr$ and hence the rotation speed $V_\mathrm{rot}$ should increase rapidly at $r\to r_\mathrm{out}$.
However, the rotation curve in figure \ref{fig:pv} (or figure \ref{fig:rot}) shows the relationship $V_\mathrm{rot} \propto r^{{-0.14}}$ and does not increase at $r_\mathrm{out}$.
{Hence, the} observational results indicate that assumption (2) is satisfied for this disk.
We used Mestel's surface density profile (3).
Under the assumptions, the first and second terms of the right side of equation (\ref{eq15}) vanish. Since the function of equation (\ref{eq3}) at $v_\mathrm{out}=r/a_\mathrm{out}<r_\mathrm{out}/a_\mathrm{out}\ll1$ is $\bm{K}(v_\mathrm{out})=\bm{K}(\sim0)=\pi/2$,
  \begin{equation}
    \frac{d\psi_\mathrm{disk}}{dr}
    =2\pi G\Sigma_\mathrm{out}\frac{a_\mathrm{out}}{r}.
    \label{eq16}
  \end{equation}
Inserting equation (\ref{eq16}) into equation (\ref{eq1}),
  \begin{equation}
    V_\mathrm{rot}^2=\frac{GM_\mathrm{BH}}{r}+2\pi G\Sigma_\mathrm{out}a_\mathrm{out},
    \label{eq17}
  \end{equation}
or
  \begin{equation}
    V_\mathrm{rot}=\sqrt{\frac{GM_\mathrm{BH}}{r}+2\pi G\Sigma_\mathrm{out}a_\mathrm{out}}.
    \label{eq18}
  \end{equation}
{If we define $\omega=r/r_\mathrm{out}$ and $\mu(\omega)\equiv r_\mathrm{out} V_\mathrm{rot}^2\omega/G$
{and use equation (\ref{eq:mdisk})},
then equation (\ref{eq17}) is
\begin{equation}
\mu(\omega)=M_\mathrm{BH}+M_\mathrm{disk}(<r_\mathrm{out})\omega  \label{eq19}
\end{equation}
which corresponds to equation (\ref{eq18}) of \citet{Hure2011}.}

  \begin{figure}[h]
  \begin{center}
     \includegraphics[width=80mm]{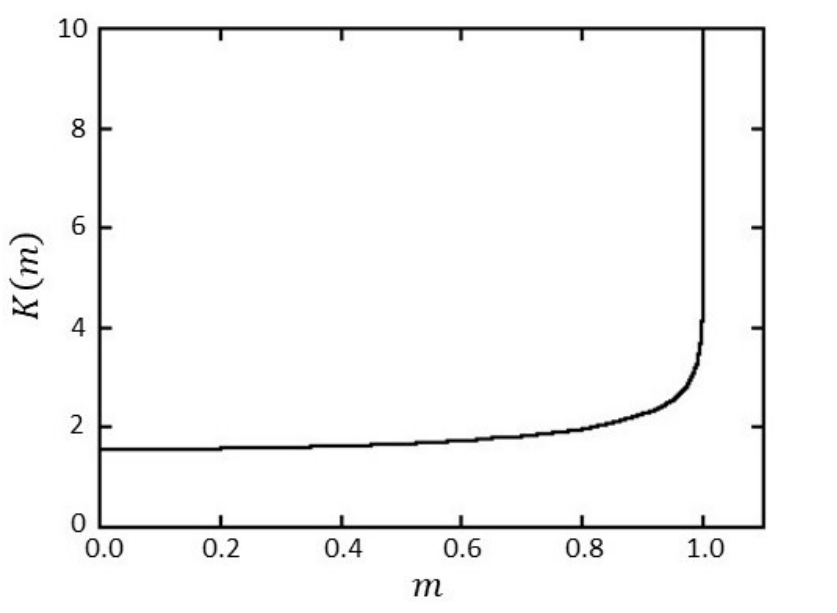}
  \end{center}
  \caption{
    \textrm{The complete elliptic integral of the first kind of equation (\ref{eq3}),
    $\bm{K}(m)$, as a function of $m$.
    As $m$ approaches 1.0, $\bm{K}(m)$ becomes extremely large.
}
      }
    \label{fig1}
  \end{figure}


\end{document}